\documentclass[onecolumn]{rsauthor}

\newcommand\apjl{Ap. J. Lett.}

\newcommand\apj{Ap. J.}
\newcommand\mnras{Mon. Not. RAS}
\newcommand\nat{Nature}
\newcommand\physrep{Phys. Rep.}

\def\lesssim{\mathrel{\hbox{\rlap{\hbox{\lower4pt\hbox{$\sim$}}}\hbox{$<$}}}}
\def\gtrsim{\mathrel{\hbox{\rlap{\hbox{\lower4pt\hbox{$\sim$}}}\hbox{$>$}}}}
\jname{Proc. R. Soc. A}

\artnum{XXXXXXXX}

\begin{document}

\title{The Origin of Long, Short and Low-luminosity Gamma-Ray Bursts}

\author{Tsvi Piran$^{1}$\thanks{* Talk given by Tsvi Piran}, Omer Bromberg$^{1}$,
Ehud Nakar$^{2}$ and Re'em Sari$^{1}$}

\address{$^1$Racah Institute of Physics, The Hebrew University, Jerusalem 91904, Israel \\
$^{2}$ Raymond and Beverly Sackler School of Physics \& Astronomy, Tel Aviv University,
Tel Aviv 69978, Israel\\}

\abstract{The origin of Gamma-Ray Bursts  is one of the most interesting puzzles in recent astronomy.
During the last decade a consensus formed that long GRBs (LGRBs) arise from the collapse of massive stars and that  short GRBs (SGRBs) have a different origin, most likely  neutron star mergers. A key ingredient of the Collapsar
model that explains how the collapse of massive stars produces a GRB is the emergence of a relativistic jet that
penetrates the stellar envelope. The condition that the emerging jet penetrates the envelope poses strong constraints on the system. %\cite{Bromberg1}.
Using these constraints  we show that: (i) Low luminosity GRBs ({\it ll}GRBs), a sub population of GRBs with a very low luminosities (and other
peculiar properties: single peaked, smooth and soft) cannot be formed by Collapsars. %\cite{Bromberg2}. 
{\it ll}GRBs must  have a different origin (most likely a shock breakout). (ii) On the other hand regular LGRBs must be formed by Collapsars. %\cite{Bromberg3}. 
(iii) While for BATSE the dividing duration  between Collapsars and non-Collapsar is indeed at $\sim 2$ sec,  the dividing duration  is different for other GRBs detectors. %\cite{Bromberg4}. 
In particular most {\it Swift} bursts longer than $0.8$ sec are of a Collapsar origin. This last results requires a revision of many conclusions concerning the origin of {\it Swift} SGRBs which were based on the commonly used  2 sec limit.
}

\date{Received xxx xxx 2012; Accepted xxx xxx 2012}

%\classification{Insert classification text here}

\maketitle

\section{Introduction}\label{sec1}

 Gamma ray bursts (GRBs) are among the most amazing transients known. In a few second a GRB emits the energy that a star like our sun emits in its whole life time. Their origin has puzzled astronomers since their serendipitous discovery in the late sixties.   After two decades in which it was believed that the GRBs are Galactic
it was realized, in the early nineties, that they have a cosmological origin \cite{Meegan+92,P92,MaoPac92}. The distance scale set immediately the energy scale to be $\gtrsim 10^{51}$ erg (including beaming corrections that were realized towards the end of the nineties \cite{Rhoads97,Rhoads99,SPH99}). Together with the short time scale this has led inevitably to the conclusion that the events involve the formation of a newborn compact  object, most likely a black hole. This conclusion has left practically just two progenitor candidates: A collapsing   massive star  or the merger of  two neutron stars (or a neutron star and a black hole).

The  observations of a few (long) GRB afterglows in 1997 revealed that those bursts arose in star forming regions. Paczynski (1998) who noticed that,  quickly  suggested that long GRBs (LGRBs) are related to core collapse events. At roughly the same time MacFadyen  \& Woosley (1999) suggested the Collapsar model. {
According to this model a GRB is produced by a relativistic jet that emerges from the center of a massive collapsing star and penetrates the stellar envelope. By now the name Collapsar is used with different variations\footnote{In some cases the term Collapsar is used generically for any model that involve a collapsing star, regardless whether there is a jet or not. In other cases  it is used more restrictively  to imply a situation in which  the collapsing star produces an accreting black hole as a central engine that drives a relativistic jet.}. We stress that here we use this original definition for a Collapsar:} {\it A  jet that penetrates the  envelope of  a Collapsing star.} Using numerical simulations MacFadyen  \& Woosley (1999)
demonstrated that a  relativistic jet can indeed penetrate a stellar envelope. Again roughly at the same time  Galama et al. (1998)  discovered that  GRB 980425 was associated with the powerful type Ic supernova: SN 1998bw. However GRB 940425 was a strange GRB. It was very weak, with an energy a few orders of magnitude less than the energy of a typical GRB. Additionally it was single peaked and smooth and it had a   very soft spectrum. It was not clear that this was a regular GRB and hence the association of GRB 980425 with SN 1998bw  was not sufficient to demonstrate a GRB-SNe association. Shortly after that Bloom et al. (1999) and others discovered red bumps in the afterglows of more distant aregular GRBs. These red bumps were interpreted as the signatures of 1998bw-like SNe, supporting the GRB-SNe association. However the evidence for a GRB-SNe association was inconclusive until  SN 2003dh was discovered in association with the regular GRB 030329 \cite{Hjorth+03,Stanek+03}. Since then a few other GRB-SNe associations were discovered. Even though most of these GRB-SNe associations are  with weak, smooth, single peaked GRBs\footnote{Another association of  a regular regular GRB and an SN,   GRB 101219B and SN 2010ma, was discovered recently.} this is generally considered as a ``proof'' of the Collapsar model for LGRBs.

An inspection of BATSE's GRBs' temporal distribution revealed \cite{Kouveliotou+93}  two groups: short ($T_{90} < 2$ sec) and long ($T_{90} > 2$ sec.).  Already in 1995
it was pointed out  \cite{CP95,P96} that the two groups have a different spatial distribution. The observed short GRBs (SGRBs) are significantly nearer (and weaker). This suggested the possibility of different physical origin for the two populations. As it takes time (and energy) to cross the relatively large stellar envelope it was argued that SGRBs cannot be produced by  Collapsars \cite{Matzner03}.   In most cases  Collapsars produce LGRBs, but  by now we know that in some cases Collapsars  produce SGRBs (see \S \ref{sec:Collapsar} below).  However a variant on this original argument that we discuss  here (in \S \ref{sec:llGRB} and \ref{sec:Collapsar} below) shows that most SGRBs cannot be produced by Collapsars.
Lack of detection of SGRB afterglows
left the situation inconclusive until  2005, when  {\it Swift} localized the first short bursts and  the first SGRBs' afterglows   were detected. It turned out that  SGRBs are not associated with star forming regions (some  arise in elliptical galaxies) and as such they are not associated with deaths of massive stars. The progenitors could be neutron star mergers (as suggested already in 1989 \cite{Eichler+89}). However as yet there is no conclusive demonstration of this origin \cite{Nakar07}.

We describe here new results, derived by Bromberg et al. (2011a, 2011b, 2012a,2012b)  concerning  the nature of GRB progenitors. We briefly discuss,   in \S \ref{sec:jet}, some recent analytic results concerning relativistic jet penetration through the stellar envelope \cite{Bromberg1}. We then consider  their implications on this picture. In \S \ref{sec:llGRB} we demonstrate that {\it}GRBs, those that appear in most GRB-SNe associations, cannot be produced by Collapsars \cite{Bromberg2}. While this weakens the case for the association of regular LGRBs with SNe, we show in \S \ref{sec:Collapsar} that  when combined with the GRBs' temporal distribution these considerations demonstrate that the LGRBs originate  from Collapsars \cite{Bromberg3}, providing a direct observational indication for jets that puncture the stellar envelope.  Further inspection of the temporal distribution enables us  to estimate (in \S \ref{sec:SGRBs}), for the first time, the fraction of Collapsars among SGRBs as a function of the observed duration \cite{Bromberg4}. We show that this fraction depends strongly on the detector (in particular on its spectral window). In particular   the standard limit of 2 sec is invalid for {\it Swift}'s observations, for which a limit of 0.8 sec is much more appropriate. 

\section{Jet Propagation}
\label{sec:jet}

\begin{figure}[!h]
%\vskip -0.5cm
\centering{\includegraphics[width=90mm]{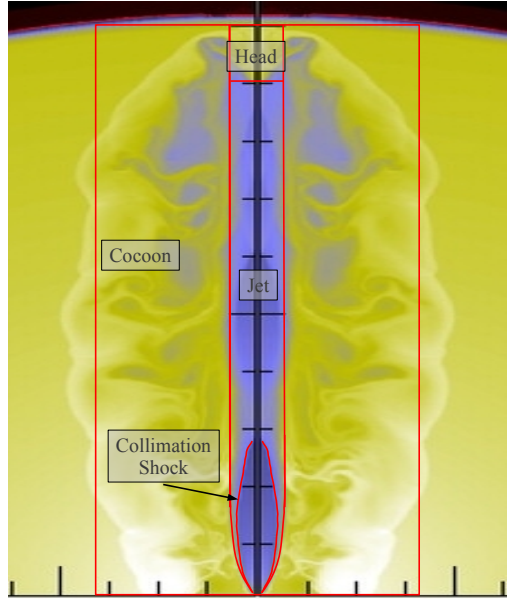}}
\vskip -1cm
\caption{ A schematic description of a jet propagating in a stellar atmosphere superimposed on a numerical jet simulation  of \cite{Morsony+07}.
%, From \cite{Bromberg1}. (Online version in colour.)}
\label{fig:Jet}}
 \end{figure}

A schematic picture of a relativistic jet propagating within a stellar envelope is depicted in Fig. \ref{fig:Jet}.  There are a few critical components. A a double shock system  appears at the head of the jet  \cite{Matzner03}. While the jet is highly relativistic these shocks slow down the head and it typically propagates with a sub or mildly relativistic velocity.  The hot material that streams sideways out of the jet's head produces a cocoon that engulfs the jet.
While expanding sideways into the rest of the stellar envelope (this expansion will eventually blow out a fraction of the stellar envelope) it also squeezes the jet and produces a (radiative)  collimation shock within the jet \cite{BrombergLevinson}.

As long as the jet is within the stellar atmosphere all its energy is dissipated at the jet's head. The total dissipated energy equals therefore the jet's luminosity times the time it takes to cross the envelope. Since the inner engine is much smaller than the envelope it is decoupled from the jet that crosses the envelope on a much larger scale  and one can expect that the luminosity before and after the jet breaks out are comparable. Using the observed GRB luminosity  to estimate the jet power before breakout we can estimate the duration of the dissipation phase as \cite{Bromberg1}:
%\begin{equation}\label{eq:tB_GRB}
%t_B \simeq 15 ~sec\cdot \epsilon_\gamma^{1/3}L_{\gamma,50}^{-1/3}\theta_{10^\circ}^{2/3}R_{11}^{2/3}M_{15\odot}^{1/3},
%\end{equation}
%where $\theta_0$, is the opening angle of the jet, $L_{\gamma}= \epsilon_\gamma L_j [{2}/({1-cos\theta_0})] $ and $\epsilon_\gamma$ is the radiative %efficiency and $A_x\equiv A/10^{x}$ in c.g.s. units.
\begin{equation}\label{eq:tB_GRB}
t_B \simeq15~{\rm sec} \cdot \left( \frac{ L_{iso}} {10^{51} {\rm~
erg/sec}}\right)^{-1/3} \left (\frac{\theta}{10^\circ}\right)^{2/3}
\left (  \frac{R_{*}} {5 R_\odot}\right)^{2/3} \left
(\frac{M_{*}}{15M_\odot}\right)^{1/3},
\end{equation}
where $ L_{iso}$ is the isotropic equivalent jet luminosity,
$\theta$ is the jet half opening angle and we have used typical
values for a LGRB. $ R_{*}$ and $M_{*}$ are the radius and the
mass of the progenitor star, where we have normalized their value
according to the typical radius and mass inferred from observations
of the few SNe associated with LGRBs.
For the jet to break out, the central engine must continue operating for a duration longer than $t_B$.  If the inner engine stops before the jet's head crosses the envelope the jet  won't produce a regular GRB.

\section{ Low luminosity GRBs}
\label{sec:llGRB}

\begin{figure}[!h]
\includegraphics[width=120mm]{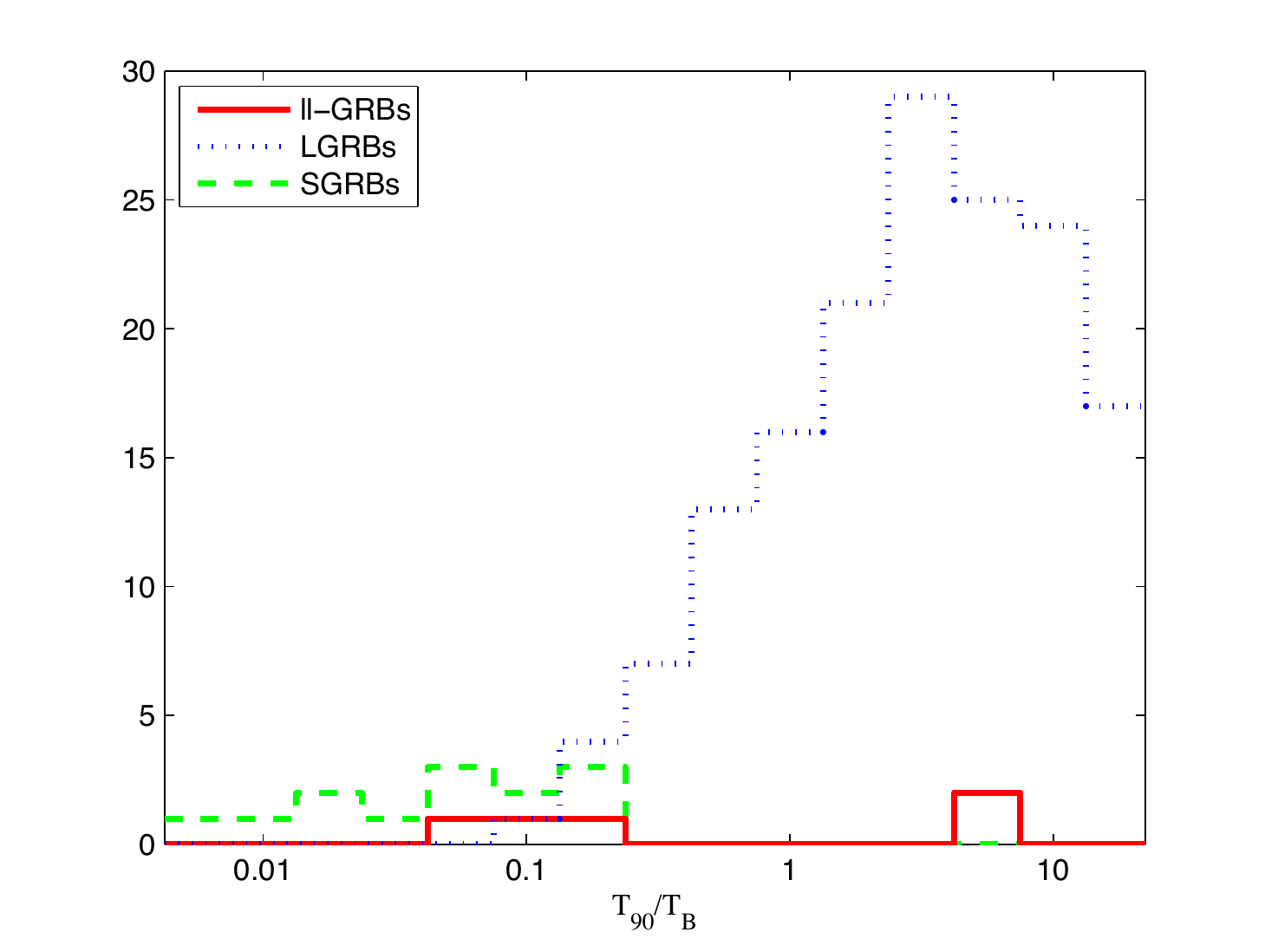}
  \caption{The distribution of $T_{90}/t_B$ for LGRBs,
  {\it ll}GRBs and SGRBs (from \cite{Bromberg2}).
 % Our samples contains the four known {\it ll}GRBs plus {\it Swift} LGRBs with redshifts.
 }
\label{fig:T90_tB}
\end{figure}

The duration of the prompt emission, approximated by  $T_{90}$, is given simply by:
\begin{equation}
T_{90} = t_{e} -  t_B  ,
\label{eq:diff}
\end{equation}
where $t_{e}$ is the total time that the engine powering the jet is active.
Within the Collapsar model,  without fine tuning,  only a small fractions of the bursts should have $T_{90}/ t_B \ll 1$ (see \S \ref{sec:Collapsar}). Namely, it is
unlikely that  the engine operates just long enough
to let the jet break out of the star and then stops  right after
breakout.  This argument was used by Matzner (2003) to argue that Collapsars cannot produce SGRBs, for which $T_{90}/ t_B \ll 1$. This is indeed confirmed in Fig \ref{fig:T90_tB} in which the distribution of $T_{90}/ t_B$ is shown for both LGRBs and SGRBs. One
can clearly see two distinct populations LGRBs for the majority of which $T_{90}/ t_B > 1$ and SGRBs, all of  which satisfy $T_{90}/ t_B <1$. 

Fig \ref{fig:T90_tB} depicts also  a third  group of GRBs, low luminosity ({\it ll}GRBs).
Like SGRBs, the observed duration distribution of {\it ll}GRBs  is inconsistent
with the predictions of the collapsar model. In particular  a  large
fraction of {\it ll}GRBs have $T_{90}/t_B\ll 1$.  The probability that
the observed {\it ll}GRBs $T_{90}/t_B$ distribution is consistent with
the LGRBs distribution is smaller than $5\%$ %\cite{Bromberg2} 
implying that  {\it ll}GRBs    cannot be generated by Collapsars and they must have a different origin. 

{\it ll}GRBs are a group of six GRBs whose luminosities are around $10^{47}-10^{49}$ ergs/sec, at least two orders of magnitude below the average luminosity of a typical GRB.
Remarkably {\it ll}GRBs are not characterized just by their low luminosity. They are also single peaked, smooth and soft.
{\it ll}GRBs include GRB9890425 (the first GRB detected to accompany a Supernova - 1998bw) as well as a few other GRB-SN pairs: GRB 031203/SN2003lw; GRB060218/SN2006aj; GRB100316D/SN2010bh. GRB051109B shows all the common features of {\it ll}GRBs but it lacks a reported SN. It is associated with a star forming region in a spiral galaxy at $z=0.08$ \cite{Perley06}.
 All {\it ll}GRBs are at very  low redshifts. With such low luminosities they couldn't have been detected from further out. While only six {\it ll}GRBs have been observed so far, given these distances, the {\it ll}GRBs inferred rate per unit volume is much larger than the rate of regular LGRBs \cite{Soderberg+06}. In fact this rate is so large that {\it ll}GRBs cannot be significantly beamed as even with a modest beaming corrections the  rate would have exceed the rates of their associated SNe - broad line type Ibc.

An interesting and likely possibility is that
 {\it ll}GRBs' jets are weak and fail to break out from their progenitors. A``failed jet'' dissipates all
its energy into the surrounding cocoon and drives its expansion. As
the cocoon reaches the edge of the star its forward shock may become
mildly or even ultra relativistic emitting the  $\gamma$-rays observed in {\it ll}GRBs
when it breaks out. This idea that {\it ll}GRBs arise from
shock breakouts  was suggested shortly following the observations of GRB980425/SN1998bw
\cite{Kulkarni+98,MacFdyen+01,Tan+01}. It drew much more attention
following the observation of additional {\it ll}GRBs with similar
properties and especially with the observation of a thermal
component in the spectrum of {{\it ll}GRB} 060218 \cite{Campana+06,Wang+07,Waxman+07}. Yet, it was hard to explain how
shock breakout releases enough energy in the form of $\gamma$-rays.
Katz, Budnik \& Waxman (2010) realized that the deviation of the breakout
radiation from thermal equilibrium provides a natural explanation to
the observed $\gamma$-rays. More recently, Nakar \& Sari (2012) calculated
the emission from mildly and ultra-relativistic shock breakouts,
including the post breakout dynamics and gas-radiation coupling.
They find that the total energy, spectral peak and duration of all
{\it ll}GRBs can be well explained  by relativistic shock
breakouts. Moreover, they find that such breakouts must satisfy a
specific relation between the observed total energy, spectral peak
and duration, and that all observed {\it ll}GRBs satisfy this relation. These
results lend a strong support to the idea that {\it ll}GRBs are
relativistic shock breakouts. From a historical point of view this
understanding closes the loop with Colgate's
(1968)  original idea, that preceded the detection of
GRBs, that  a SN shock breakout will produce a GRB.

As we discuss in the following section, the observed GRB duration distribution indicates the existence
of many "failed jets" in which the engine time is shorter than the breakout time. This is consistent with the
observations that the rate of {\it ll}GRBs is much higher than the rate of regular LGRBs.

\section{Long GRBs and Collapsars}
\label{sec:Collapsar}

As most of the GRBs associated with SNe are {\it ll}GRBs    one might think at first that this new understanding rules out the Collapsar model for LGRBs. However, on the contrary, these arguments provide a new and unexpected direct observational confirmation of the Collapsar origin of  LGRBs.
%\cite{Bromberg3}.  
Consider again Eq. \ref{eq:diff}. Under very general conditions this equation results in a flat duration distribution  for durations significantly shorter than the typical breakout time.

It follows from Eq. \ref{eq:diff} that the  distribution, $p_\gamma(T_{90})$ of  the observed GRB durations is  a convolution of  $p_e(t_e)$, the distribution of  engine operating times, and $p_B(t_B)$ the distribution of  jet breakout times. Under quite general conditions (more specifically unless $p_e$ varies very rapidly around $t_B$, an unlikely situation)  the following limits  hold:
%\cite{Bromberg3}:
\begin{equation}\label{tg}
p_\gamma (T_{90})  \approx \left\{
\begin{array}{cc}
  p_e(t_B) &  T_{90} \lesssim t_B   \\
%  F(t_\gamma) dt_\gamma&  t_\gamma \sim t_b   \\
  p_e(T_{90}) &   T_{90} \gg t_B
\end{array}
\right. .
\end{equation}

\begin{figure}[!h]
%\vskip -2cm
\includegraphics[width=120mm]{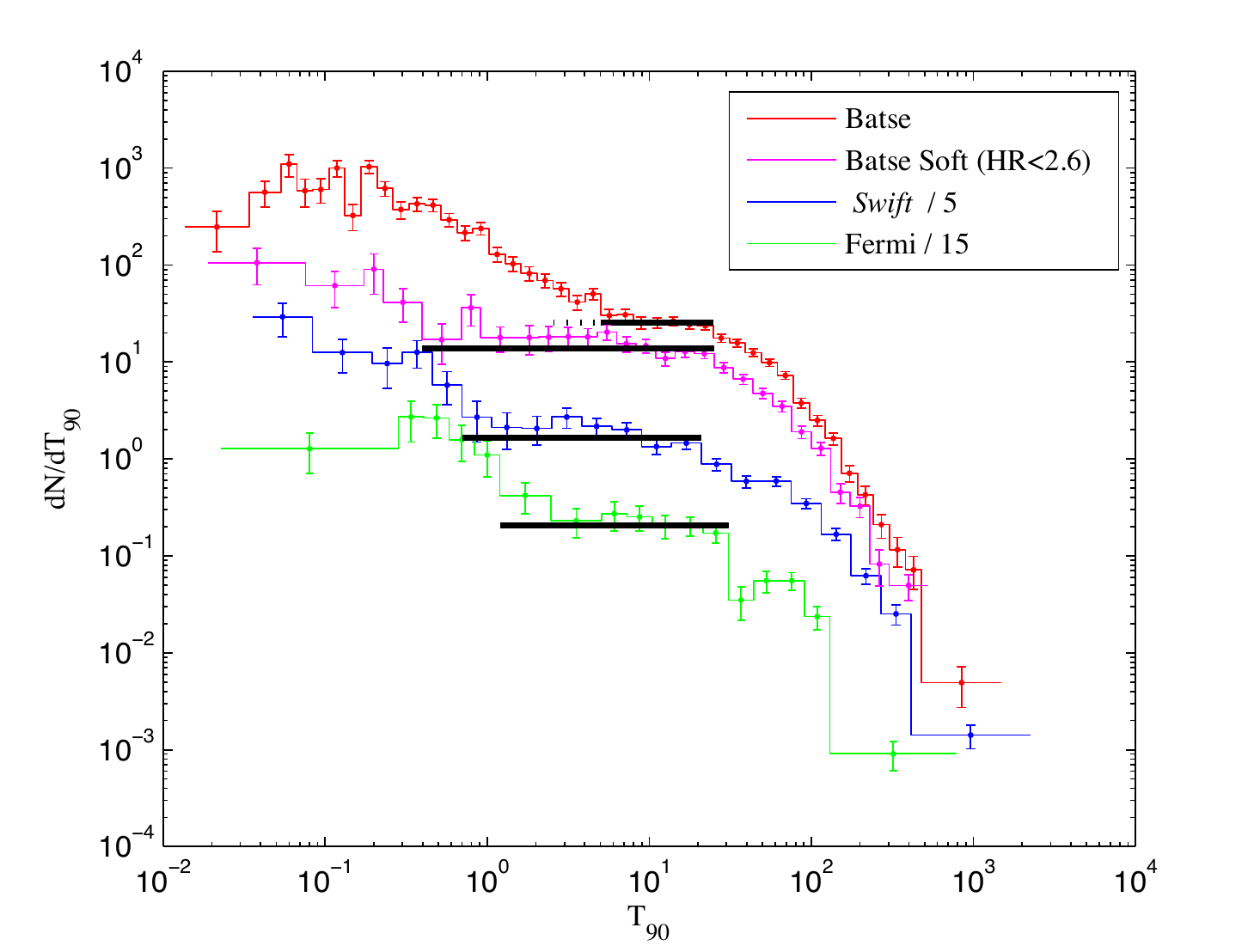}
  \caption{The duration  distributions, $dN/dT_{90}$, of BATSE (red), {\it Swift} (blue) and Fermi GBM (green) GRBs.
    Also plotted is the distribution of the soft (hardness ratio $< 2.6$) BATSE bursts (magenta).
    For clarity the {\it Swift} values are divided by a factor of 5 and the Fermi GBM by 15.
    Note that the quantity $dN/dT$  is depicted and
    not ${dN}/{d\log T}$ as traditionally shown in such plots \cite{Kouveliotou+93}.
    The black lines show the best fitted flat interval in each data set:
    $5-25$ sec (BATSE), $0.7-21$ sec ({\it Swift}),  and $2.5-31$ sec (Fermi). The upper limits of this range
    indicate a typical breakout time of a few
    dozens seconds, in agreement with the prediction of the Collapsar model.
    Soft BATSE bursts  show a considerably longer plateau
    ($0.4-25$ sec), indicating that most of the soft short bursts are in fact
    Collapsars (from \cite{Bromberg3}.)}
\label{fig:T-dist}
\end{figure}

Particularly interesting for our purpose here is the appearance at short durations $T_{90} \lesssim t_B$ of a  flat region, a plateau, in which the rate of events is independent of the duration. Remarkably such plateaus exist in all the observed GRB duration distributions (see Fig. \ref{fig:T-dist}). They weren't noticed so far because the "canonical" distribution plot \cite{Kouveliotou+93} depicts $d N/d\log(T)$ instead of $dN/dT$. 

The  higher end  of the plateau enables us to estimate $t_B$ and from this to infer some the basic properties of the collapsing stars. We find, for example,  that a typical  progenitor size is $ \sim 5 R_\odot$.  
Another interesting feature seen in Fig. \ref{fig:T-dist}  is the rapid decline at durations longerr than $t_B$. At this regime
according to Eq. \ref{tg} the distribution is dominated by $p_e(t_e)$, thus $p_e(T_{90})\approx p_{\gamma}(T_{90})$.  An extrapolation  of this distribution to shorter engine operating times suggests that there are numerous cases in which $t_e < t_B$ and the jet fails to break out. This is in a very nice agreement with the very large inferred event rate of  {\it ll}GRB if these are interpreted as``failed jets''.

The appearance of these plateau (and their dependence on the observed hardness) is the first direct observational confirmation of the Collapsar model. More specifically these plateau demonstrates the fact that the duration of an LGRB is  the difference between  two independent time scales. This confirms a basic prediction of the Collapsar model:  the overall duration is the difference between the time that the engine operates and the time it takes the jet to penetrate the stellar envelope.

At very short durations the plateau doesn't extend all the way to zero. This does not rule out the model. At this regime non-Collapsar SGRBs, that have a different origin, appear and dominate the distribution. As different detectors have different relative sensitivities to long (and soft) vs. short (and hard) GRBs the duration at which  short non-Collapsars  begin to dominate varies from one detector to another\footnote{In principle we should have worked with a redshift corrected sample. However such samples of SGRBs are too small and we are forced to use  the observed durations. This imply that we have to worry also about time dilation which, in turn depend on the sensitivity of the detector and the corresponding depth of the samples. However, as SGRBs are detected only from relatively small redshifts this time dilation correction is not significant.}.  To demonstrate this dependence on the detector's spectral window we artificially change BATSE's effectiveness for detection of  hard SGRBs by considering  only softer BATSE bursts (hardness ratio $< 2.6$). As expected,  for this  softer BATSE sample the non-Collapsar peak shrinks and the plateau extends down  to shorter durations.

\section{Short non - Collapsar  GRBs}
\label{sec:SGRBs}

Shortly after Kouveliotou et al. (1993) demonstrated that there are two populations of GRBs: long and short ones it became clear that these two populations have  different spatial distributions and a different origin \cite{CP95,P96}. By now we know that 
LGRBs arise from Collapsars.  SGRBs  have other projenitors, most likely neutron star mergers. As we are still uncertain concerning the origin of SGRBs we  denote them here as non-Collapsars. So far it was implicitly assumed that the division line between long and SGRB is  at 2 sec regardless of the observing satellite. The existence of a plateau in the observed LGRBs (of Collapsar origin) duration distribution enables us to determine, for the first time, the fractions of Collapsars vs. non-Collapsars as a function of the observed time for every specific detector. While we cannot determine if a specific bursts is a Collapsar or not we can give now %\cite{Bromberg4} 
a probabilistic estimate for a given duration and hardness.

The basic idea is very simple. For a given detector we determine the rate of detection of bursts within the plateau. { This provides an estimate for the  detection rate
of short duration Collapsars by this detector}. Now we can compare this rate to the rate of SGRBs at any given duration and obtain the Collapsar and non-Collapsar fractions as a function of duration. This estimate is performed  for different detectors or even for different detection windows (hardness) for a specific detector.

\begin{figure}[!h]
\includegraphics[width=120mm]{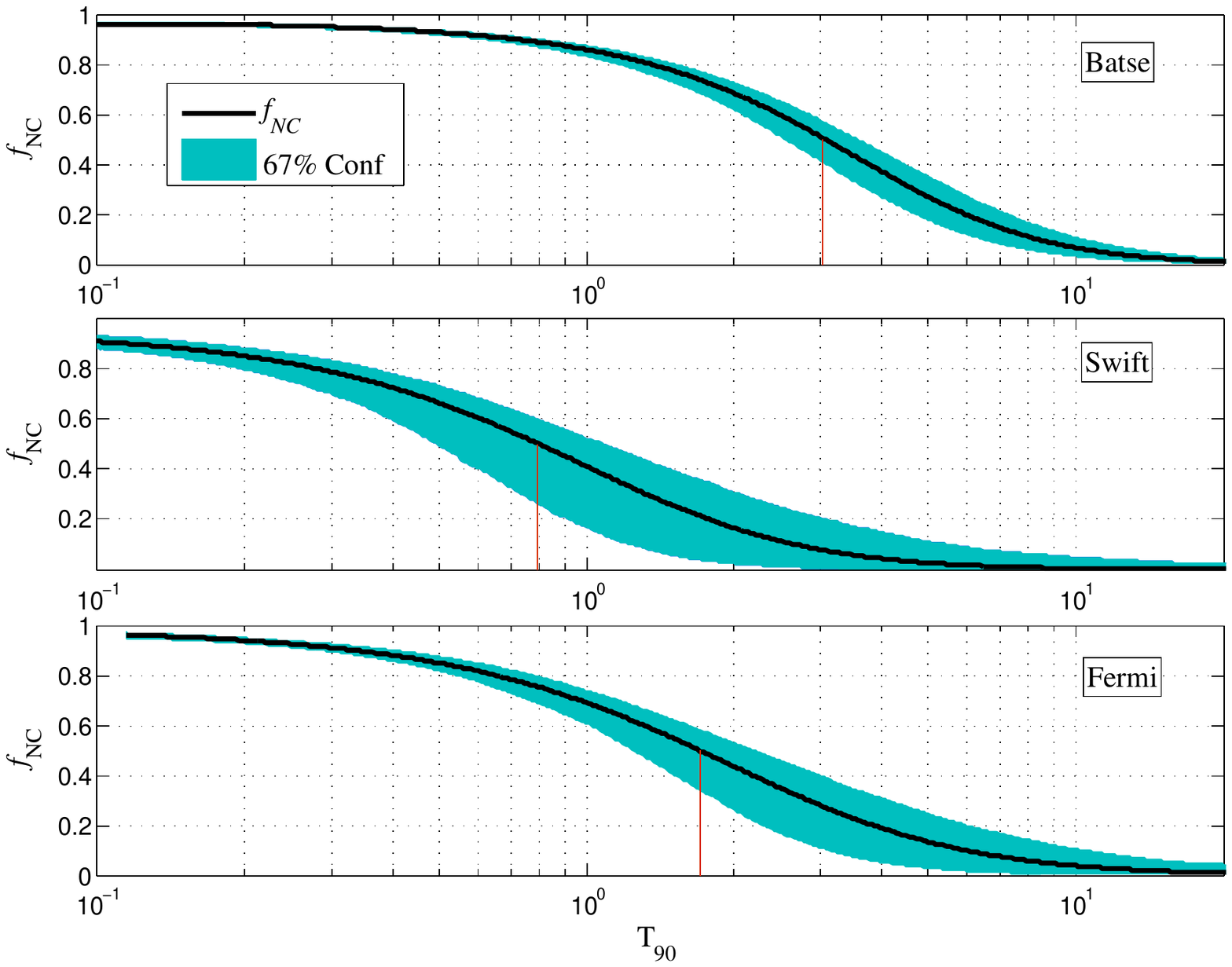}
  \caption{ From top to bottom: BATSE, {\it Swift} and
Fermi GBM  fractions, $f_{NC}$ of non-Collapsars from the total number of observed GRBs as a
function of the observed duration, $T_{90}$.  The shaded regions represent 67\% confidence limits of $f_{NC}$.
Also plotted {in red} are the $T_{90}$ values for which $f_{NC} = 0.5$  (from \cite{Bromberg4}).}
\label{fig:Short_fraction}
\end{figure}

{
We have fitted the different duration distributions with   a plateau (representing  Collapsars)  and a lognormal distribution (for  non-Collapsars). The fit is remarkably good and it enables us to estimate the fraction, $f_{NC}$ of non-Collapsars from the total number of observed GRBs as a
function of the observed duration, $T_{90}$ (see Fig. \ref{fig:Short_fraction}).
For Batse   $T_{90} < 2$ sec is a reasonable  threshold to identify non-Collapsars.
This limit results in a probability $> 70\%$ for a correct classification for  BATSE bursts.
However, this condition is  misleading for  {\it Swift} bursts.
{At $T_{90}=2$ sec a {\it Swift} burst has a $84{\pm14}\%$ probability to be a Collapsar!
Clearly,  for {\it Swift}  a $2$ sec division line results in a large number of misidentified Collapsars as non-Collapsars.
We propose to draw the division line between collapsars and non-Collapsars at the duration where
the probability that a GRB is a non-Collapsar is 50\%.
With this condition, a BATSE GRB can be classified as a non-Collapsar if its $T_{90}<3.1{\pm0.5}$ sec.
{\it Swift} bursts can be identified as non-collapsars only if their duration $T_{90} < 0.8{\pm 0.3}$ sec,
while the corresponding limit for GBM  is  $T_{90} < 1.7^{+0.5}_{-0.5}$ sec.}
The results shown here can be expanded and improved when we consider the hardness of the bursts.
}

Using these results one can go back and examine various studies of short bursts that attempted to compare 
various features of short (standing for non-Collapsars) with long (Collapsars) and check the samples used.
A preliminary inspection of such studies %\cite{Bromberg4} 
reveals that while in some cases the short sample has only a small number of potential Collapsars in other  cases the short
sample used was heavily contaminated by high probability potential Collapsars (with observed duration shorter than 2 sec) and it is possible and even likely that these have dominated the results.  { Note in particular that our results are not inconsistent with those of Berger (2011), who finds  a larger dispersion in properties of {\it Swift} SGRBs (with $T_{20}< 2 $ sec host galaxies and the positions of the bursts within the hosts, as compared with the more homogenous properties of LGRBs host and their position within the hosts. A mixed sample that contains both Collapsars and non-Collapsars is expected to show such larger dispersion (see \cite{Bromberg4} for further details). }

\section{Conclusions}

To conclude we summarize our basic findings. We define\footnote{Note that the term Collapsars has also been used in both wider and narrower contents.} a Collapsar as a collapsing massive star that produces in it center  a relativistic jet. The jet  penetrates the stellar envelope and produces the GRB ones it has left the star. The duration of the envelope penetration phase depends on the jet's luminosity, its opening angle and the  size and density profile of the stellar envelope.

One can expect that the duration of a burst is typically comparable to or longer than the jet breakout time. Indeed, a  comparison of the estimated jet breakout time of a typical LGRB shows that it is shorter than the observed duration of the burst. On the other hand the jet breakout time is much
longer than the duration of a short burst. This provides the first indication that SGRBs are not produced by Collapsars. We have shown that
a third group  of low luminosity GRBs  also don't satisfy this condition. This implies that {\it ll}GRBs don't arise from Collapsars.

Within the Collapsar model  the observed duration of a GRB is the difference between the time its central engine  operates, producing the jet, and the jet's breakout time. This directly implies that at short durations the rate of bursts produced by Collapsars
should be independent of their duration. We have shown that such a behavior is observed in the duration distributions of all GRB satellites: BATSE, {\it Swift}
and GBM.  This provides the a direct observational confirmation  of the basic prediction of the  Collapsar  model and as such demonstrates the Collapsar origin of LGRBs.

This last feature also enables us to determine the fraction of Collapsars within the observed SGRBs. This fraction depends on the characteristics of the detector. For BATSE the standard division between Collapsar and non-Collapsars is indeed at $\sim 2 $ sec. However for the softer {\it Swift}
many bursts shorter than 2 sec are of Collapsar origin. This might have led to some confusion in the past in interpreting observations of these short bursts as indications for properties of non-Collapsar GRBs.

\vskip 1cm
This research was supported by an Advanced ERC
 grant and by the Israeli center for Excellence for
High Energy AstroPhysics  (TP), by an ERC starting grant and an ISF grant  (EN) and by
Packard, Guggenheim and Radcliffe fellowships (RS).

\end{document}